\documentclass[12pt]{article}
\usepackage{bbold,amssymb,hyperref,graphicx,enumerate}

\newcommand{\be}{\begin{equation}}
\newcommand{\ee}{\end{equation}}
\newcommand{\bea}{\begin{eqnarray}}
\newcommand{\eea}{\end{eqnarray}}
\newcommand{\beas}{\begin{eqnarray*}}
\newcommand{\eeas}{\end{eqnarray*}}

\begin{document}
\begin{titlepage}

\begin{flushright}
{\small OU-HET-1142}
\end{flushright}

\medskip

\begin{center}

{\Large Extractable entanglement from a Euclidean hourglass}

\vspace{12mm}

\renewcommand\thefootnote{\mbox{$\fnsymbol{footnote}$}}
Takanori Anegawa${}^{1}$\footnote{takanegawa@gmail.com},
Norihiro Iizuka${}^{1}$\footnote{iizuka@phys.sci.osaka-u.ac.jp},
Daniel Kabat${}^{2,3}$\footnote{daniel.kabat@lehman.cuny.edu}

\vspace{6mm}

${}^1${\small \sl Department of Physics, Osaka University} \\
{\small \sl Toyonaka, Osaka 560-0043, JAPAN}

\vspace{3mm}

${}^2${\small \sl Department of Physics and Astronomy} \\
{\small \sl Lehman College, City University of New York} \\
{\small \sl 250 Bedford Park Blvd.\ W, Bronx NY 10468, USA}

\vspace{3mm}

${}^3${\small \sl Graduate School and University Center, City University of New York} \\
{\small \sl  365 Fifth Avenue, New York NY 10016, USA}

\end{center}

\vspace{12mm}

\noindent
We previously proposed that entanglement across a planar surface can be obtained from the partition function on a Euclidean hourglass geometry.
Here we extend the prescription to spherical entangling surfaces in conformal field theory.  We use the prescription to evaluate log terms in the entropy of a CFT in two dimensions,
a conformally-coupled scalar in four dimensions, and a Maxwell field in four dimensions.  For Maxwell we reproduce the extractable entropy obtained
by Soni and Trivedi.  We take this as evidence that the hourglass prescription provides a Euclidean technique for evaluating extractable entropy in quantum field theory.

\end{titlepage}
\setcounter{footnote}{0}
\renewcommand\thefootnote{\mbox{\arabic{footnote}}}

\hrule
\tableofcontents
\bigskip
\hrule

\addtolength{\parskip}{8pt}
\section{Introduction}
A prescription for defining entanglement entropy was introduced in \cite{Anegawa:2021osi}, building on an earlier proposal by Solodukhin \cite{Solodukhin:2004rv,Solodukhin:2005qy}.
The main motivation was to discuss entanglement between two regions $A$ and $\bar{A}$ {\em without} attempting to tensor-factor the Hilbert space into
${\cal H}_A \otimes {\cal H}_{\bar{A}}$.  Instead a single density matrix $\rho_\epsilon$ was introduced, defined on the entire Hilbert space.  The parameter $\epsilon$ is a UV regulator that
cuts off the large quantum fluctuations present near the entangling surface.  Such a regulator is necessary to make $\rho_\epsilon$ a well-defined density matrix on the entire Hilbert space.

Our proposal is that regulated entanglement entropy should be defined as half of the von Neumann entropy of $\rho_\epsilon$.
As a heuristic motivation for this proposal, in the limit $\epsilon \rightarrow 0$ the density matrix $\rho_\epsilon$ formally approaches the tensor product of the reduced
density matrices associated with regions $A$ and $\bar{A}$.
\be
\rho_\epsilon \stackrel{\epsilon \rightarrow 0}{\longrightarrow} \rho_A \otimes \rho_{\bar{A}}
\ee
This is only a heuristic motivation since the subregion density matricies $\rho_A$, $\rho_{\bar{A}}$ are not well-defined.  The proposal to work in terms of $\rho_\epsilon$
avoids any need to tensor-factor the Hilbert space and is therefore well-defined in continuum quantum field theory.  It is manifestly gauge invariant and has a direct geometric
interpretation as evaluating the partition function on a Euclidean ``hourglass'' geometry.

In a seemingly unrelated line of development, several authors have shown that standard
replica methods when applied to Maxwell theory do not give the correct (quantum) entanglement entropy \cite{Casini:2015dsg,Soni:2016ogt,Casini:2019nmu}.  That is, they do not count the number of Bell pairs split by the
entangling surface.  This was clarified by Soni and Trivedi \cite{Soni:2016ogt}, who argued that replica methods include a classical or Shannon contribution to the entropy associated with classical correlations across the entangling surface required by the Gauss constraint.  Soni and Trivedi were able to correct the replica result, subtracting the Shannon contribution to obtain what they referred to as
extractable entropy.  They obtained the log divergence in the extractable entropy for a spherical entangling surface in four dimensions, reproducing a coefficient previously obtained by other
authors \cite{Kamenshchik:1993kh,Dowker:2010bu,Casini:2015dsg}.  The mismatch between replica methods (which lead to anomaly coefficients) and the coefficient of the log term in the entanglement entropy of a Maxwell field has been further clarified by Casini, Huerta, Mag\'an and Pontello \cite {Casini:2019nmu}.

The present paper has three main goals.
\begin{enumerate}
\item
Our previous work \cite{Anegawa:2021osi} considered planar entangling surfaces and evaluated the entropy in a variety of simple theories, both conformal and non-conformal.  Here we specialize to conformal field theories and extend the prescription to spherical entangling surfaces.
\item
As a warm-up we evaluate the log divergence in the entropy for a general conformal field theory in 2D and for a conformally-coupled scalar field in 4D.
\item
We evaluate the log divergence in the entropy for a Maxwell field in 4D and show that the coefficient of the log agrees with the extractable entropy obtained by
Soni and Trivedi \cite{Soni:2016ogt}.
\end{enumerate}
We take this agreement as strong evidence that the hourglass prescription provides a direct geometric method for calculating extractable entropy in quantum field theory.
Using a slightly different method fields with spin on a hyperbolic cylinder were considered by David and Mukherjee \cite{David:2020mls}, who showed that this approach
reproduces the log coefficient in the extractable entropy even for gravity \cite{David:2022jfd}.

An outline of this paper is as follows.  We develop the hourglass prescription for a 2D CFT on a spatial circle in section \ref{sect:2D}, where the entangling surface
consists of two points.  In section \ref{sect:higher} we extend the prescription to spherical regions in conformal field theories in higher dimensions.  We illustrate the prescription in four dimensions
in sections \ref{sect:scalar} and \ref{sect:Maxwell} by calculating the log divergence in the entropy for a conformally-coupled scalar field and for a Maxwell field.  In the latter case we find agreement with the extractable entropy
computed by Soni and Trivedi.  We conclude in section \ref{sect:conclusions}.

\section{Hourglass prescription in 2D\label{sect:2D}}
\begin{figure}
\centerline{\includegraphics[width=7cm]{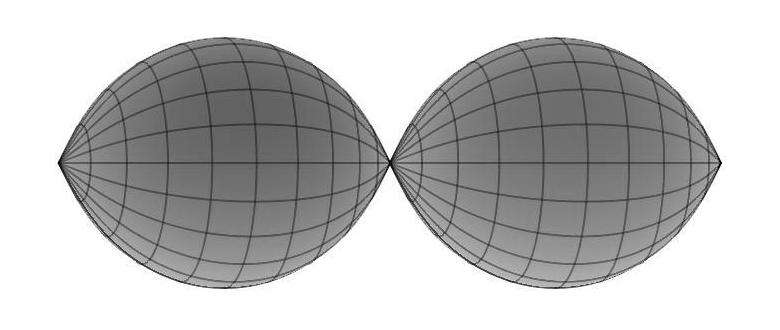} \hspace{2mm} \raisebox{1.4cm}{$\Rightarrow$} \hspace{2mm} \includegraphics[width=7cm]{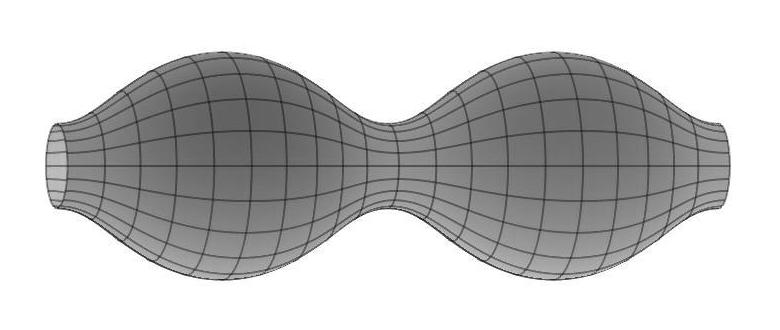}}
\caption{Two touching footballs, smoothed out into an hourglass geometry.  The two football tips that don't touch should be identified.  Likewise the two open ends of the hourglass
should be identified.\label{fig:hourglass}}
\end{figure}

We begin by considering a 2D CFT on a unit spatial circle and make a division of the circle into $A \cup \bar{A}$ where
\be
\bar{A} = \lbrace -\pi < \phi < 0 \rbrace \qquad A = \lbrace 0 < \phi < \pi \rbrace
\ee
To study entanglement between $A$ and $\bar{A}$ we introduce the operator
\be
\label{Vdef}
V = \int_{-\pi}^\pi d\phi \, \vert \sin \phi \, \vert \, T_{00}(\phi)
\ee
The operator $V$ is singular on the boundary between the two regions (meaning at $\phi = 0,\,\pi$) where $\vert \sin \phi \, \vert$ isn't smooth,
but this is easy to regulate.  We introduce a parameter $\epsilon \rightarrow 0$ and define
\be
\label{Vepsdef}
V_\epsilon = \int_{-\pi}^\pi d\phi \, \sqrt{\sin^2 \phi + \epsilon^2} \, T_{00}(\phi)
\ee
We define a partition function by
\be
\label{Zeb}
Z_\epsilon(\beta) = {\rm Tr} \, e^{-\beta V_\epsilon}
\ee
and propose to define a regulated entanglement entropy by
\be
\label{Sdef}
S_\epsilon = {1 \over 2} \left.\left(\beta {\partial \over \partial \beta} - 1\right)\right\vert_{\beta = 2\pi} \left(-\log Z_\epsilon\right)
\ee
The regulator function we have introduced $\sqrt{\sin^2 \phi + \epsilon^2}$ is a convenient explicit choice but it is not unique.  Any function which is smooth and non-zero near
$\phi = 0,\,\pi$ would do equally well.  The partition function (\ref{Zeb}) corresponds to putting the theory on a Euclidean geometry\footnote{We explain this connection in more detail in appendix \ref{appendix:Hamiltonian}, where we show that $V_\epsilon$ is the Hamiltonian that generates translations of the Euclidean time
coordinate $\theta$.}
\bea
\label{2Dhourglass}
&& ds^2 = d\phi^2 + \left(\sin^2 \phi + \epsilon^2\right) d\theta^2 \\[3pt]
\nonumber
&& \phi \approx \phi + 2 \pi \qquad \theta \approx \theta + \beta
\eea
When $\epsilon = 0$ the geometry looks like two spheres (two American footballs if $\beta \not= 2\pi$) touching at their tips.  The regulator smooths the geometry into an ``hourglass'' shape that is topologically a torus, as shown in Fig.\ \ref{fig:hourglass}.

Having presented our proposal we should give some motivation and connect it to discussions in the literature.
To do this we start from the modular Hamiltonian appropriate to the division into $A \cup \bar{A}$ \cite{Cardy:2016fqc}.
\be
K = \int_{-\pi}^\pi d\phi \, \sin \phi \, T_{00}(\phi)
\ee
In the ground state $K$ is formally related to the reduced density matrices for regions $A$ and $\bar{A}$ by $e^{-2 \pi K} = \rho_A \otimes \rho_{\bar{A}}^{-1}$.
This is only a formal relation in continuum field theory, since the Hilbert space does not admit a tensor factorization, and for that reason the following motivation is purely heuristic.
Given the absolute value we introduced in (\ref{Vdef}), which amounts to a sign change in region $\bar{A}$, we see that $V$ is formally related to the reduced density matrices by
\be
e^{-2 \pi V} = \rho_A \otimes \rho_{\bar{A}}
\ee
If we define a partition function
\be
Z(\beta) = {\rm Tr} \, e^{-\beta V}
\ee
then formally the entanglement entropy is given by
\be
\label{S}
S = {1 \over 2} \left.\left(\beta {\partial \over \partial \beta} - 1\right)\right\vert_{\beta = 2\pi} \left(-\log Z\right)
\ee
The factor of ${1 \over 2}$ compensates for the overcounting of having two regions, so that (\ref{S}) gives the von Neumann entropy for just one of the reduced density matrices.
This provides a heuristic motivation for the proposal (\ref{Sdef}).

Two comments regarding this prescription are in order.
\begin{enumerate}
\item
From a geometric perspective, note that the hourglass has a freely-acting Killing vector ${\partial \over \partial \theta}$.  In this sense it is similar to an ordinary thermal system in the imaginary-time formalism.
\item
From a canonical perspective the partition function (\ref{Zeb}) corresponds to a density matrix
\be
\rho = {1 \over Z_\epsilon(\beta)} e^{-\beta V_\epsilon}
\ee
This can be thought of as a thermal density matrix with a position-dependent proper temperature
\be
T_{\rm proper} = {1 \over \beta \sqrt{\sin^2 \phi + \epsilon^2}}
\ee
Note that the density matrix is defined on the entire Hilbert space.  Thus we're able to discuss entanglement without introducing a tensor factorization of the Hilbert space, thereby avoiding a problematic issue in
continuum quantum field theory \cite{Witten:2018lha}.
\end{enumerate}
For completeness we evaluate the hourglass entropy in a general 2D CFT.  With a change of coordinates $d\phi = \sqrt{\sin^2\phi + \epsilon^2} \, d\chi$ the metric becomes conformal
to a flat torus.
\be
ds^2 = \left(\sin^2\phi + \epsilon^2\right) \left(d\chi^2 + d\theta^2\right)
\ee
We can neglect the conformal factor since it doesn't contribute to the entropy \cite{deAlwis:1994ej}.  The periodicity has changed, from $\phi \approx \phi + 2 \pi$ to
$\chi \approx \chi + L_\epsilon$ where
\be
\label{Lepsilon}
L_\epsilon = \int_0^{2\pi} {d\phi \over \sqrt{\sin^2 \phi + \epsilon^2}} = 4 \log {1 \over \epsilon} + {\rm finite}
\ee
On a long cylinder the partition function per unit length is $- \log Z / L_\epsilon = - {\pi c \over 6 \beta}$ \cite{Bloete:1986qm}, so the log divergence in the entropy is given by
\bea
\nonumber
S & = & {1 \over 2} \left.\left(\beta {\partial \over \partial \beta} - 1\right)\right\vert_{\beta = 2 \pi} \left(- {\pi c L_\epsilon \over 6 \beta}\right) \\
& = & {c \over 3} \log {1 \over \epsilon}
\eea
Thus we recover the standard result for the entanglement of an interval in 2D CFT \cite{Holzhey:1994we,Calabrese:2004eu}.

\section{Hourglass prescription in higher dimensions\label{sect:higher}}
In what follows we'll denote the number of spacetime dimensions by $d$ and (less frequently) the number of spatial dimensions by $n$.

To generalize the hourglass prescription to higher dimensions we begin from the modular Hamiltonian for a spherical region in conformal field theory.
As shown by \cite{Casini:2011kv} and reviewed in appendix \ref{appendix:modular},
this leads us to the Euclidean de Sitter metric written in static coordinates.
\bea
\label{sphere}
&&ds^2_{\rm sphere} = R^2\left[d\phi^2 + \sin^2 \phi \, d\theta^2 + \cos^2 \phi \, d\Omega_{d-2}^2\right] \\[3pt]
\nonumber
&&0 \leq \phi \leq \pi/2 \qquad \theta \approx \theta + 2 \pi
\eea
This describes a round sphere of radius $R$, as one can see by parametrizing the hypersurface $\lbrace u^2 + v^2 + \vert \vec{w} \vert^2 = 1\rbrace \in {\mathbb R}^{d+1}$ as
\bea
\nonumber
&& u = \sin \phi \cos \theta \\
&& v = \sin \phi \sin \theta \\
\nonumber
&& \vec{w} = \cos \phi \, \vec{n} \qquad \hbox{\rm with $\vec{n} \in {\mathbb R}^{d-1}$, $\vert \vec{n} \vert = 1$}
\eea
To obtain a singular football geometry analogous to (\ref{2Dhourglass}) we extend the range of $\phi$ to $-\pi/2 \leq \phi \leq \pi/2$ and modify the periodicity to $\theta \approx \theta + \beta$.
\bea
\label{football}
&&ds^2_{\rm football} = R^2\left[d\phi^2 + \sin^2 \phi \, d\theta^2 + \cos^2 \phi \, d\Omega_{d-2}^2\right] \\[3pt]
\nonumber
&&-\pi/2 \leq \phi \leq \pi/2 \qquad \theta \approx \theta + \beta
\eea
One can think of this as two Euclidean de Sitter spaces at inverse temperature $\beta$ that touch at their common horizon, that is, at the $S^{d-2}$ located at $\phi = 0$.  Finally to smooth
the geometry into an hourglass we take
\bea
\label{hourglass}
&&ds^2_{\rm hourglass} = R^2 d\phi^2 + \left(R^2 \sin^2 \phi + \epsilon^2\right) d\theta^2 + R^2 \cos^2 \phi \, d\Omega_{d-2}^2 \\[3pt]
\nonumber
&&-\pi/2 \leq \phi \leq \pi/2 \qquad \theta \approx \theta + \beta
\eea
Here $\epsilon \rightarrow 0$ is a dimensionful regulator with units of length.  We recognize the first and last terms in (\ref{hourglass}) as the metric on a round
$(d-1)$-sphere,\footnote{In ${\mathbb R}^d$ with coordinates $(\vec{x},z)$ set $\vec{x} = \cos \phi \, \vec{n}$ and $z = \sin \phi$.  Here $-{\pi \over 2} \leq \phi \leq {\pi \over 2}$ and $\vert \vec{n} \vert$
is a unit vector.} so
\bea
\label{hourglass2}
&&ds^2_{\rm hourglass} = \left(R^2 \sin^2 \phi + \epsilon^2\right) d\theta^2 + R^2 \, d\Omega_{d-1}^2 \\[3pt]
\nonumber
&& \theta \approx \theta + \beta
\eea
Regarding $\theta$ as the Euclidean time direction, this describes a spherical space of radius $R$ with a temperature that depends on the azimuthal angle $\phi$.
\be
T_{\rm proper} = {1 \over \beta \sqrt{R^2 \sin^2 \phi + \epsilon^2}}
\ee
The temperature is lowest at the poles $\phi = \pm \pi / 2$ (the centers of the static patches) and highest at the equator $\phi = 0$ (the common de Sitter horizon).

For calculational purposes it's convenient to switch to a different conformal frame in which the proper temperature is constant.  This is the so-called optical geometry of \cite{Gibbons:1976pt}, applied in this context in \cite{Barbon:1994ej,Emparan:1994qa,deAlwis:1994ej}.\footnote{The change of frame corresponds to a change in integration measure which, as discussed in \cite{deAlwis:1994ej}, produces an anomalous term in the effective action that doesn't affect the entropy.  Intuitively the change in integration
measure, being local, shifts the effective action by a term proportional to $\beta$ and hence does not affect the entropy.}
\bea
\nonumber
ds^2_{\rm optical} & = & {1 \over g_{\theta\theta}} ds^2_{\rm hourglass} \\
\label{optical}
& = & d\theta^2 + {1 \over \sin^2 \phi + (\epsilon/R)^2} \underbrace{\left(d\phi^2 + \cos^2 \phi d\Omega_{d-2}^2\right)}_{d\Omega^2_{d-1}} \\
\nonumber
& & \hspace{-2cm} \theta \approx \theta + \beta \qquad - {\pi \over 2} \leq \phi \leq {\pi \over 2}
\eea
The proper temperature is now $1/\beta$ everywhere but the spatial geometry is no longer a round $S^{d-1}$.  Instead it can be thought of as two copies of hyperbolic space ${\cal H}^{d-1}$, cut off at large radius and
smoothly attached to each other.  To see this we change coordinates on the northern hemisphere and set $\sinh \rho = {1 / \tan \phi}$.
The northern hemisphere ${\pi \over 2} \geq \phi > 0$ corresponds to $0 \leq \rho < \infty$.
This puts the metric on the northern hemisphere in the form
\be
\label{optical2}
ds^2_{\rm optical} = d\theta^2 + {1 \over 1 + ({\epsilon \over R})^2 \cosh^2\rho} \underbrace{\left(d\rho^2 + \sinh^2 \rho d\Omega_{d-2}^2\right)}_{ds^2_{{\cal H}^{d-1}}}
\ee
We recognize $d\rho^2 + \sinh^2 \rho \, d\Omega_{d-2}^2$ as the metric on a unit ${\cal H}^{d-1}$.  Thus we can approximate the spatial geometry as
two copies of hyperbolic space with unit radius of curvature that are
\begin{enumerate}[(i)]
\item
cut off at a radial coordinate $\rho_0 \sim \log {2 R \over \epsilon}$, and
\item
connected by a neck region which, as can be seen from (\ref{optical}), is approximately a strip $- {\epsilon \over R} < \phi < {\epsilon \over R}$ around the equator of a round sphere of radius $R/\epsilon$.
\end{enumerate}

Thus for a conformal field theory in any number of dimensions we are instructed to compute a thermal partition function $Z_\epsilon(\beta)$ on the spatial geometry (\ref{optical2}).  Entanglement entropy is then given by a formula
analogous to (\ref{Sdef}).
\be
\label{Sdef2}
S_\epsilon = {1 \over 2} \left.\left(\beta {\partial \over \partial \beta} - 1\right)\right\vert_{\beta = 2\pi} \left(-\log Z_\epsilon\right)
\ee
Although the prescription applies in any number of dimensions, in the examples that follow we specialize to $d = 4$ and focus on obtaining the log divergent terms in the entropy.

\section{Conformal scalars in 4D\label{sect:scalar}}
In this section we consider a massless conformally-coupled scalar field in the optical geometry (\ref{optical}).  We specialize to four spacetime dimensions, $d = 4$.
Our goal is to calculate the partition function on this space as $\epsilon \rightarrow 0$.  There is a leading quadratic divergence $\sim R^2 / \epsilon^2$ that depends on the choice of
regulator function in (\ref{hourglass}) which we will largely ignore.  Instead we are interested in keeping track of the subleading $\sim \log (R/\epsilon)$ divergence since it is universal.

We begin by studying the spatial 3-geometry in more detail.  One quantity of interest is the spatial volume, which can be evaluated in terms of elliptic integrals and expanded for small $\epsilon$.\footnote{Conventions differ.
Here $E(k) = \int_0^1 dx \, \sqrt{1 - k^2 x^2 \over 1 - x^2}$.}
\bea
\nonumber
{\rm vol}_3 & = & 4\pi \int_{-\pi/2}^{\pi/2} d\phi \, {\cos^2 \phi \over \big(\sin^2 \phi + (\epsilon/R)^2\big)^{3/2}} \\[5pt]
\nonumber
& = &  {8\pi R^2 \over \epsilon^2} \Bigg[{\sqrt{R^2 + \epsilon^2} \over R} E\Big({R \over \sqrt{R^2 + \epsilon^2}}\Big) - {\epsilon^2 \over R \sqrt{R^2 + \epsilon^2}} K\Big({R \over \sqrt{R^2 + \epsilon^2}}\Big) \Bigg] \\[5pt]
\label{OpticalVolume}
& = & {8 \pi R^2 \over \epsilon^2} - 4 \pi \log {R \over \epsilon} + {\rm finite}
\eea
As expected the volume diverges as $\epsilon \rightarrow 0$.  In general the quadratic divergence gets contributions from both the hyperbolic and neck regions of the geometry, which
is another way of saying that it's sensitive to the choice of regulator function.  But we're particularly interested in the log divergence, and it's important to recognize that the log divergence
{\em only comes from the hyperbolic part of the geometry.}  As a direct test of this, consider the volume of hyperbolic space with a radial cutoff at $\rho_0 \sim \log {R \over \epsilon}$.
\bea
\nonumber
&& ds^2_{{\cal H}^3} = d\rho^2 + \sinh^2 \rho \, d\Omega_2^2 \\
&& {\rm vol}({\rho < \rho_0}) = \int_0^{\rho_0} d\rho \, 4 \pi \sinh^2 \rho = {\pi R^2 \over 2 \epsilon^2} - 2 \pi \log {R \over \epsilon} + {\rm finite}
\eea
Multiplying by 2 to account for the two copies of ${\cal H}^3$, the coefficient of the log divergence agrees with (\ref{OpticalVolume}).

We will also be interested in the curvature of the spatial geometry.  The optical 3-geometry is conformal to a sphere,
\be
\label{ds3}
ds_3^2 = \underbrace{1 \over \sin^2 \phi + (\epsilon/R)^2}_{\Omega^2} \underbrace{\left(d\phi^2 + \cos^2 \phi d\Omega_{d-2}^2\right)}_{ds_0^2}
\ee
where $ds_0^2$ is the metric on a unit 3-sphere.  The scalar curvature can be obtained from the conformal transformation rule
\be
{\cal R} = {1 \over \Omega^2} {\cal R}_0 - {2(n-1) \over \Omega^3} \Box_0 \Omega - {(n-1)(n-4) \over \Omega^4} g_0^{ab} \partial_a \Omega \partial_b \Omega
\ee
Here $n = 3$ is the number of spatial dimensions and ${\cal R}_0 = n (n-1)$ is the scalar curvature of $S^n$.  This leads to
\bea
{\cal R} & = & {-6 \sin^2\phi + 4 (\epsilon/R)^2 \cos^2\phi + 6 (\epsilon/R)^4 \over \sin^2\phi + (\epsilon/R)^2} \\[5pt]
\nonumber
& \approx & \left\lbrace\begin{array}{ll} -6 & \quad \hbox{\rm if $\vert \phi \vert > {\epsilon \over R}$} \\[5pt] +4 & \quad \hbox{\rm if $- {\epsilon \over R} < \phi < {\epsilon \over R}$} \end{array}\right.
\eea
This is the scalar curvature in three dimensions, but since the 4-geometry is metrically a product with $S^1$ it's also the scalar curvature in
four dimensions.  Note that the scalar curvature is bounded everywhere.  In the regions that can be approximated by hyperbolic space we have ${\cal R} \approx -6$ as
expected.\footnote{The scalar curvature of a unit ${\cal H}^{n}$ is ${\cal R} = - n(n-1)$.}  It appears that the curvature varies rapidly near $\phi = 0$, but this is a
coordinate artifact since the proper length of the interval $- {\epsilon \over R} < \phi < {\epsilon \over R}$ is ${\cal O}(1)$.

Thus we're led to a picture where the partition function (\ref{Zeb}) is a standard thermal partition function, evaluated at inverse temperature $\beta$, on a spatial geometry which has a diverging
volume but ${\cal O}(1)$ curvature as $\epsilon \rightarrow 0$.  We consider a massless conformally-coupled scalar on this geometry for which
\be
- \log Z = {1 \over 2} {\rm Tr} \, \log \left(-\partial_\theta^2 - \nabla_3^2 + \xi {\cal R}\right)
\ee
Here $\nabla_3^2$ is the Laplacian on the spatial geometry (\ref{ds3}) and for conformal coupling in 4D we set $\xi = {d-2 \over 4(d-1)} = 1/6$.  In a proper-time parametrization with UV cutoff $\Lambda \rightarrow \infty$
we have
\be
\label{HeatKernel}
- \log Z = - {1 \over 2} \int_{1/\Lambda^2}^\infty {ds \over s} \, {\rm Tr} e^{-s\left(-\partial_\theta^2 - \nabla_3^2 + \xi {\cal R}\right)}
\ee
On a product space the heat kernel in (\ref{HeatKernel}) factors.
\begin{itemize}
\item
The factor associated with the thermal circle is
\be
K_\beta(s) = \sum_{n \in {\mathbb Z}} e^{-s (2 \pi n / \beta)^2}
\ee
By Poisson resummation this can be re-expressed as a sum over winding modes.
\be
\label{Kbeta}
K_\beta(s) = {\beta \over \sqrt{4 \pi s}} + {\beta \over \sqrt{\pi s}} \sum_{m = 1}^\infty e^{-m^2 \beta^2 / 4s}
\ee
The first term in (\ref{Kbeta}) makes a contribution to $-\log Z$ which is UV divergent but proportional to $\beta$.  Such a term doesn't contribute to the entropy so we will discard it.
The remaining terms are all UV finite since the heat kernel provides a UV cutoff at $s \sim \beta^2$.
\item
The factor associated with the optical 3-geometry is
\bea
\label{K3}
&& K_3(s) = \int d^3x \sqrt{g_3} \, K_3(s,x,x) \\
\nonumber
&& K_3(s,x,x') = \langle x \vert e^{-s\left(- \nabla_3^2 + \xi {\cal R}\right)} \vert x' \rangle
\eea
This appears difficult to evaluate, but recall that the optical geometry has a curvature that is ${\cal O}(1)$.  We therefore expect that the heat kernel $K_3(s,x,x')$ has a finite limit as
$\epsilon \rightarrow 0$.  There is still a divergence in (\ref{K3}) due to the infinite volume of optical space, but recall that the log divergence in (\ref{OpticalVolume}) only comes from the hyperbolic part
of the geometry.  So to extract the log divergence we replace $K_3$ with the heat kernel on hyperbolic space.\footnote{This approximation is adequate to capture log divergences
in the final answer.  There are also quadratic divergences coming from the neck region of the geometry (\ref{optical2}) which we will not attempt to calculate.  Since the neck region
enters as a UV cutoff in the original conformal frame (\ref{hourglass}) it will not contribute to a log divergence.}  In appendix \ref{appendix:KH} we show that at coincident points $x = x'$
this replacement gives
\be
\label{replacement}
K_3(s,x,x) \rightarrow {1 \over (4 \pi s)^{3/2}}
\ee
Rather remarkably this is the same result that one would obtain for a massless field in flat space.
\end{itemize}

Now it's a simple matter of assembling the pieces.  We have
\bea
\nonumber
- \log Z(\beta) & = & - {\beta \over \sqrt{4 \pi}} \int_0^\infty {ds \over s^{3/2}} \sum_{m = 1}^\infty e^{-m^2 \beta^2 / 4 s} \int d^3x \sqrt{g_3} \, K_3(s,x,x) \\
\label{logZscalar}
& \rightarrow & - {\beta \over \sqrt{4 \pi}} \int_0^\infty {ds \over s^{3/2}} \sum_{m = 1}^\infty e^{-m^2 \beta^2 / 4s} \, {\rm vol}_3 \, {1 \over (4 \pi s)^{3/2}} \\
\nonumber
& = & - \sum_{m = 1}^\infty {1 \over \pi^2 \beta^3 m^4} \, {\rm vol}_3 \\
\nonumber
& = & - {\pi^2 \over 90 \beta^3} \, {\rm vol}_3
\eea
The entropy is then
\be
S = {1 \over 2} \left.\left(\beta {\partial \over \partial \beta} - 1\right)\right\vert_{\beta = 2 \pi} (- \log Z) = {1 \over 360 \pi} \, {\rm vol}_3
\ee
Recalling the expression for the volume (\ref{OpticalVolume}) we have
\be
S =  \# \, {R^2 \over \epsilon^2} - {1 \over 90} \log {R \over \epsilon} + {\rm finite}
\ee
The coefficient of the quadratic divergence is not universal and is not determined by this calculation, but the coefficient of the log divergence is trustworthy.
It agrees with a previous result due to Dowker \cite{Dowker:2010bu}, in which entanglement entropy was obtained from thermodynamics in de Sitter space.
It also matches the partition function on a de Sitter instanton evaluated by Kamenshchik \cite{Kamenshchik:1993kh}.

\section{Maxwell field in 4D\label{sect:Maxwell}}
Finally we consider a Maxwell field in four spacetime dimensions.  Entanglement in Maxwell theory has been the subject of a long series of works \cite{Donnelly:2011hn,Eling:2013aqa,Casini:2013rba,Casini:2014aia,Donnelly:2014fua,Huang:2014pfa,Ghosh:2015iwa,Aoki:2015bsa,Donnelly:2015hxa,Soni:2015yga,Casini:2015dsg,Soni:2016ogt,Agarwal:2016cir,Aoki:2017ntc,Huerta:2018xvl,Casini:2019nmu,Huerta:2022cqw}.
Here we take advantage of conformal symmetry and calculate the log divergence in the entropy of a spherical region.  The coefficient of the log agrees with
a previous calculation by Dowker \cite{Dowker:2010bu} and also agrees with the log term in the extractable entropy evaluated by Soni and Trivedi \cite{Soni:2016ogt}.  We take this as strong
evidence that the hourglass prescription provides a direct geometric way of computing physical (extractable) entanglement entropy in quantum field theory.

Our starting point is the gauge-fixed action
\bea
\nonumber
S & = & \int d^4x \, \sqrt{g} \left( {1 \over 4} F_{\mu\nu} F^{\mu\nu}  + {1 \over 2} (\nabla_\mu A^\mu)^2 - i b \nabla_4^2 c \right) \\
& = & \int d^4x \, \sqrt{g} \left( {1 \over 2} A^\mu \left(-g_{\mu\nu} \nabla_4^2 + R_{\mu\nu}\right) A^\nu - i b \nabla_4^2 c \right)
\eea
We are working in Euclidean space in Feynman gauge.  The ghost fields $b$, $c$ behave as minimally-coupled scalars while
the gauge field $A_\mu$ couples to the Ricci curvature $R_{\mu\nu}$.  The four dimensional Laplacian
$\nabla_4^2 = \nabla_\mu \nabla^\mu$ acts in the appropriate representation, either spin-0 or spin-1.

It's convenient to decompose the metric and Laplacian as
\bea
\nonumber
ds^2 & = & d\theta^2 + ds_3^2 \\
\nabla_4^2 & = & \partial_\theta^2 + \nabla_3^2
\eea
We likewise decompose $A_\mu = (A_\theta,\,A_i)$ into a Euclidean time component $A_\theta$ and spatial components $A_i$.
These behave as a (scalar, vector) from the 3D point of view.  In a proper-time parametrization we have
\be
- \log Z = - {1 \over  2} \int_{1/\Lambda^2}^\infty {ds \over s} \left( {\rm Tr} \, e^{-s(-\nabla_4^2)} + {\rm Tr} \, e^{-s(-g_{ij} \nabla_4^2 + R_{ij})} \right)  + \int_{1/\Lambda^2}^\infty {ds \over s} \, {\rm Tr} \, e^{-s (-\nabla_4^2)}
\ee
from $A_\theta$, $A_i$ and the ghosts, respectively.  (For $A_i$ note that $\nabla_4^2$ acts in the spin-1 representation.)  There's a partial
cancellation between $A_\theta$ and the ghosts, so we're left with
\bea
\nonumber
- \log Z & = & - {1 \over  2} \int_{1/\Lambda^2}^\infty {ds \over s} \left( - {\rm Tr} \, e^{-s(-\nabla_4^2)} + {\rm Tr} \, e^{-s(-g_{ij} \nabla_4^2 + R_{ij})} \right) \\[5pt]
& = & - {1 \over 2} \int_0^\infty {ds \over s} K_\beta(s) \left( - {\rm Tr} \, e^{-s(-\nabla_3^2)} + {\rm Tr} \, e^{-s(-g_{ij} \nabla_3^2 + R_{ij})} \right)
\eea
In the second line we factored out the heat kernel for the thermal circle (\ref{Kbeta}), dropping the term with no winding since it doesn't contribute to the entropy.
This let us remove the UV cutoff $\Lambda$ from the calculation.

From the spatial point of view we have a massless minimally-coupled scalar and a massless vector.  The scalar heat kernel is given in (\ref{MasslessScalarK}), while for the
vector heat kernel we borrow the result from \cite{Giombi:2008vd}.
\bea
\label{HeatKernels}
&& {\rm Tr} \, e^{-s(-\nabla_3^2)} = {\rm vol}_3 \, {e^{-s} \over (4 \pi s)^{3/2}} \\
\nonumber
&& {\rm Tr} \, e^{-s(-g_{ij} \nabla_3^2 + R_{ij})} = {\rm vol}_3 \, {e^{-s} + 2 + 4s \over (4 \pi s)^{3/2}}
\eea
There's an amusing cancellation and we're left with
\bea
\label{Zmaxwell}
- \log Z(\beta) & = & - {\beta \over \sqrt{4 \pi}} \int_0^\infty {ds \over s^{3/2}} \sum_{m = 1}^\infty e^{-m^2 \beta^2 / 4s} \, {\rm vol}_3 \, {2 + 4s \over (4 \pi s)^{3/2}} \\
\nonumber
& = & - {1 \over \pi^2} \sum_{m = 1}^\infty \left({1 \over \beta m^2} + {2 \over \beta^3 m^4} \right) \, {\rm vol}_3 \\
\nonumber
& = & - \left( {1 \over 6 \beta} + {\pi^2 \over 45 \beta^3} \right) \, {\rm vol}_3
\eea
The entropy is then
\be
S = {1 \over 2} \left.\left(\beta {\partial \over \partial \beta} - 1\right)\right\vert_{\beta = 2 \pi} (- \log Z) = {4 \over 45 \pi} \, {\rm vol}_3
\ee
Recalling the expression for the volume (\ref{OpticalVolume}) we have
\be
S =  \# \, {R^2 \over \epsilon^2} - {16 \over 45} \log {R \over \epsilon} + {\rm finite}
\ee
We have not determined the coefficient of the quadratic divergence.  The coefficient of the log agrees with the coefficient in the extractable entropy obtained by
Soni and Trivedi \cite{Soni:2016ogt}, who wrote their result in terms of the area of the entangling surface as $D \log {A \over \epsilon^2}$ with $D = - {16 \over 90}$.
It also agrees with the thermal entropy in de Sitter space evaluated by Dowker \cite{Dowker:2010bu} and with the partition function on a de Sitter instanton
evaluated by Kamenshchik \cite{Kamenshchik:1993kh}.

\section{Conclusions\label{sect:conclusions}}
In this paper we extended the hourglass prescription to spherical entangling surfaces in conformal field theory.  For a Maxwell field in four dimensions we
showed that the coefficient of the log divergence agrees with the coefficient in the extractable entropy obtained by Soni and Trivedi \cite{Soni:2016ogt}.
We take this as strong evidence that the hourglass prescription provides
a direct geometric method for computing extractable entanglement in field theory.  That is, the hourglass prescription
counts the number of Bell pairs split by the entangling surface.  It avoids any Shannon contribution to the entropy arising from classical correlations across
the entangling surface.  (Such correlations are present in a gauge theory due to the Gauss constraint.)  We view this as a further advantage of the hourglass prescription,
in addition to the fact that it is manifestly gauge invariant and avoids any need to tensor-factor the Hilbert space.

Let us mention a few connections with the literature and directions for future work.

\noindent
{\em Interpreting the Maxwell result} \\
Casini and Huerta \cite{Casini:2015dsg} studied the entanglement of a Maxwell field across a spherical entangling surface and showed that the Maxwell field decomposes into
two massless scalars from which the $\ell = 0$ mode has been removed.  To connect this to the present work, note that for a massless scalar in 4D (assuming conformal coupling) we can borrow the result (\ref{logZscalar}).
\be
\label{logZscalar2}
- \log Z_{\rm scalar} = - {\beta \over \sqrt{4 \pi}} \int_0^\infty {ds \over s^{3/2}} \sum_{m = 1}^\infty e^{-m^2 \beta^2 / 4s} \, {\rm vol}_3 \, {1 \over (4 \pi s)^{3/2}}
\ee
The $\ell = 0$ mode has a Dirichlet boundary condition at the origin \cite{Casini:2015dsg}.  It behaves like a field in two dimensions.  Dropping the $S^{d-2}$ in (\ref{hourglass}), we
see that it propagates on a geometry
\be
ds^2 = R^2 d\phi^2 + \left(R^2 \sin^2 \phi + \epsilon^2\right) d\theta^2
\ee
which is nothing but the two-dimensional hourglass studied in section \ref{sect:2D}.  On the 2D hourglass the Dirichlet condition at the origin (and likewise at infinity) corresponds to a
Dirichlet condition at $\phi = \pm {\pi \over 2}$.  We can use a conformal transformation to turn the hourglass into a long cylinder with Dirichlet boundary conditions at the ends, however
in place of (\ref{Lepsilon}) the length of the cylinder is
\be
\label{Lepsilon2}
{\rm vol}_1 = \int_{-\pi/2}^{\pi/2} {R \, d\phi \over \sqrt{R^2 \sin^2 \phi + \epsilon^2}} = 2 \log {R \over \epsilon} + {\rm finite}
\ee
The partition function for the $\ell = 0$ mode can be obtained from (\ref{logZscalar2}) by
\begin{enumerate}
\item
replacing ${\rm vol}_3$ with ${\rm vol}_1$
\item
replacing $1/(4 \pi s)^{3/2}$ with
$1/(4 \pi s)^{1/2}$, the heat kernel appropriate to one spatial dimension
\end{enumerate}
This leads to
\be
\label{logZscalarleq0}
- \log Z_{\rm scalar}^{\ell = 0} = - {\beta \over \sqrt{4 \pi}} \int_0^\infty {ds \over s^{3/2}} \sum_{m = 1}^\infty e^{-m^2 \beta^2 / 4s} \, {\rm vol}_1 \, {1 \over (4 \pi s)^{1/2}}
\ee
It's then straightforward to check that\footnote{We are retaining only log terms in the volume.  For the log terms from (\ref{OpticalVolume}) we have ${\rm vol}_1 = - {1 \over 2 \pi} {\rm vol}_3$.}
\be
-2(\log Z_{\rm scalar} - \log Z_{\rm scalar}^{\ell = 0}) = - {\beta \over \sqrt{4 \pi}} \int_0^\infty {ds \over s^{3/2}} \sum_{m = 1}^\infty e^{-m^2 \beta^2 / 4s} \, {\rm vol}_3 \, {2 + 4s \over (4 \pi s)^{3/2}}
\ee
This agrees with the Maxwell result (\ref{Zmaxwell}), as Casini and Huerta predicted.

\noindent
{\em Conformal anomaly} \\
On general grounds one expects log divergences in a conformal field theory to be determined by the conformal anomaly, or equivalently by the ${\cal O}(s^0)$ terms in the expansion of the
heat kernel \cite{Birrell:1982ix}, which when integrated over a smooth manifold imply
\be
\int \sqrt{g} \, \langle T^\mu{}_\mu \rangle = a E_4 + c W^2
\ee
Here the integrated Euler density and square of the Weyl tensor are
\bea
\label{E4}
&& E_4 = {1 \over 64 \pi^2} \int \sqrt{g} \left({\rm Riemann}^2 - 4 \, {\rm Ricci}^2 + R^2\right) \\
\label{W2}
&& W^2 = - {1 \over 64 \pi^2} \int \sqrt{g} \left({\rm Riemann}^2 - 2 \, {\rm Ricci}^2 + {1 \over 3} R^2\right)
\eea
This leads to a connection between entanglement entropy (calculated using the replica trick) and anomaly coefficients \cite{Ryu:2006ef,Solodukhin:2008dh,Casini:2011kv,Eling:2013aqa}.
It would be interesting to explore how the hourglass prescription modifies this connection.  In this regard let us note that the hourglass (\ref{hourglass}) is obtained by periodically identifying along a Killing vector
${\partial \over \partial \theta}$.  This means integrated quantities such as (\ref{E4}), (\ref{W2}) are proportional to $\beta$ and therefore do not contribute to the entropy.\footnote{The Euler number can't
vary continuously, which provides a quick argument that the Euler number of the hourglass (topologically $S^1 \times S^3$) vanishes.  Note the discrete difference from Euclidean de Sitter space (topologically $S^4$) with Euler number $2$.}  However the arguments in the literature connecting entropy and anomalies only
apply when a UV regulator is introduced and held fixed while a singular limit of the geometry is taken \cite{Soni:2016ogt}.  The heat kernel on a singular hourglass geometry
may be well-defined but this is a subtle situation to analyze.  Further subtleties with anomalies have been studied in \cite{Casini:2019nmu}.

\noindent
{\em Future directions and open questions} \\
There are many interesting directions and open questions to explore.

In the examples we considered we focused on log terms in the entropy in even spacetime dimensions.  But the basic prescription (\ref{optical2}), (\ref{Sdef2}) applies to spherical regions in conformal field
theory in any number of dimensions.  It would be particularly interesting to calculate finite terms in the entropy in odd spacetime dimensions, especially in $d = 3$ as a measure of topological entanglement entropy
\cite{Kitaev:2005dm,PhysRevLett.96.110405}.

It would also be interesting to apply the hourglass prescription to spherical regions in conformal field theories that have a holographic dual.  Applying the hourglass prescription on the boundary, perhaps one could identify the bulk dual of the
calculation along the lines of \cite{Lewkowycz:2013nqa}.

More ambitiously it would be interesting to extend the prescription to non-spherical regions.  Here we face an obstacle, that very little is known about the starting point (the modular Hamiltonian) for non-spherical regions,
even in conformal field theory.  A more tractable possibility might be to extend the prescription to spherical regions in non-conformal theories.  Work in this direction is in progress \cite{AIK}.

\bigskip
\goodbreak
\centerline{\bf Acknowledgements}
\noindent
The work of TA and NI were supported in part by JSPS KAKENHI Grant Number 21J20906(TA), 18K03619(NI). 
The work of NI was also supported by MEXT KAKENHI Grant-in-Aid for Transformative Research Areas A ``Extreme Universe'' No.\ 21H05184.
DK is supported by U.S.\ National Science Foundation grant PHY-2112548.

\appendix
\section{Hourglass Hamiltonian\label{appendix:Hamiltonian}}
The two dimensional hourglass geometry (\ref{2Dhourglass}) is engineered to have a Killing vector ${\partial \over \partial \theta}$.
Moreover it is designed so that the Hamiltonian that generates shifts along the Killing vector is the operator $V_\epsilon$.  To show this we begin from the
general formula for the conserved charge associated with a Killing vector $\xi^\mu$.
\be
\label{conserved}
Q = - \int_\Sigma d^{d-1}x \sqrt{g_\Sigma} \, T^{\mu\nu} n_\mu \xi_\nu
\ee
Here $T^{\mu\nu}$ is the stress tensor, $\Sigma$ is a hypersurface with induced metric $g_\Sigma$ and $n^\mu$ is a unit vector
normal to the hypersurface.  The fact that $Q$ is conserved, and generates the transformation $x^\mu \rightarrow x^\mu + \xi^\mu$, follows from the Ward identity\footnote{A detailed
discussion may be found in \cite{osborn2019lectures,Simmons-Duffin:2016gjk}.}
\be
\nabla_\mu \langle T^{\mu\nu}(x) {\cal O}(x_1) \cdots {\cal O}(x_n) \rangle = - \sum_{i = 1}^n {1 \over \sqrt{g}} \delta^d(x - x_i) \nabla_i^\nu \langle {\cal O}(x_1) \cdots {\cal O}(x_n) \rangle
\ee
(multiply by the Killing vector and integrate).

Let's evaluate $Q$ on the hourglass geometry
\be
ds^2 = d\phi^2 + \left(\sin^2 \phi + \epsilon^2\right) d\theta^2
\ee
The Killing vector is $\xi = {\partial \over \partial\theta}$.  We take $\Sigma$ to be a hypersurface of constant $\theta$ with induced metric $ds^2_\Sigma = d\phi^2$ and unit normal
\be
n = {1 \over \sqrt{\sin^2 \phi + \epsilon^2}} \, {\partial \over \partial \theta}
\ee
Note that the Killing vector is proportional to the normal vector, $\xi^\mu = \sqrt{\sin^2 \phi + \epsilon^2} \, n^\mu$.  This lets us write the conserved charge
purely in terms of the normal vector.
\be
\label{Q}
Q = - \int d\phi \, \sqrt{\sin^2 \phi + \epsilon^2} \, T_{\mu\nu} n^\mu n^\nu
\ee
We want to compare $Q$ to the operator $V_\epsilon$ defined in (\ref{Vepsdef}).  One complication is that $V_\epsilon$ is written on the $t = 0$ slice of Minkowski space while $Q$
is written in Euclidean signature.\footnote{For further discussion of this point see section 7.1 of \cite{Simmons-Duffin:2016gjk}.}  We account for this by setting $n^\mu = -i n_{\scriptscriptstyle M}^\mu$ where $n_{\scriptscriptstyle M}^\mu = (1,0,\ldots,0)$ is a unit
vector normal to the Minkowski $t = 0$ slice.  Then in Lorentzian signature we find that
\be
Q = \int d\phi \, \sqrt{\sin^2 \phi + \epsilon^2} \, T_{00}
\ee
which agrees with $V_\epsilon$.

An analogous calculation in higher dimensions would show that the generator of $\tau$ translations on the geometry (\ref{Kgeom}) is
\be
Q = - \int d^{d-1}x \, {R^2 - \rho^2 \over 2 R^2} T_{\mu\nu} n^\mu n^\nu
\ee
Comparing this to $1/R$ times the generator $K$ defined in (\ref{Kdef}), we see that the two agree after $Q$ is continued to Lorentzian signature.

\section{Modular Hamiltonian for spherical regions\label{appendix:modular}}
In this appendix we review the steps \cite{Casini:2011kv} leading from the modular Hamiltonian for a spherical region in CFT${}_d$ to the Euclidean de Sitter metric (\ref{sphere}).

Consider dividing the $t = 0$ slice of Minkowski space into two regions separated by a sphere of radius $R$.  For a conformal field theory the modular Hamiltonian for such a division is \cite{Hislop:1981uh}
\be
\label{Kdef}
K = \int d^{d-1}x \, {R^2 - \rho^2 \over 2 R} T_{00}
\ee
where $\rho$ is a radial coordinate.  We will focus on the interior region $0 \leq \rho < R$, although in conformal field theory there is a symmetry $\rho \rightarrow R^2 / \rho$ that exchanges the interior and exterior.
The partition function $Z(\beta) = {\rm Tr} \, e^{-\beta K / R}$ corresponds to putting the theory on a Euclidean geometry
\bea
\label{Kgeom}
&& ds^2 = {(R^2 - \rho^2)^2 \over 4 R^4} d\tau^2 + d\rho^2 + \rho^2 d\Omega_{d-2}^2 \\
\nonumber
&& \tau \approx \tau + \beta
\eea
Defining the partition function in this way makes $\tau$ and $\beta$ dimensionful.  The on-shell temperature that makes the geometry smooth is $\beta = 2 \pi R$.

Consider a Weyl transformation $ds^2 \rightarrow {4 R^4 \over (R^2 + \rho^2)^2} ds^2$ together with a change of coordinates $r = {2 R^2 \rho \over \rho^2 + R^2}$.  This brings the metric to the form
\be
ds^2 = {R^2 - r^2 \over R^2} d\tau^2 + {R^2 \over R^2 - r^2} dr^2 + r^2 d\Omega_{d-2}^2
\ee
which is Euclidean de Sitter space.  Setting $\tau = R \theta$ and $r = R \cos \phi$ puts the metric in the form used in section \ref{sect:higher}.
\be
\label{sphere2}
ds^2 = R^2\left[\sin^2 \phi \, d\theta^2 + d\phi^2 + \cos^2 \phi \, d\Omega_{d-2}^2\right]
\ee
This describes a round sphere of radius $R$ when
\be
\theta \approx \theta + 2 \pi \qquad 0 \leq \phi \leq \pi/2
\ee
Note that Euclidean time $\theta$ is now dimensionless and the on-shell temperature corresponds to $\beta = 2\pi$.
In these coordinates the center of the static patch is at $\phi = \pi/2$ and the de Sitter horizon is at $\phi = 0$.
So far we've only discussed the interior region, but really the interior and exterior geometries are identical, and
as shown in section \ref{sect:higher} they can be smoothly connected once a regulator is introduced.

\section{Heat kernel on hyperbolic space\label{appendix:KH}}
Heat kernels in hyperbolic space have been been studied in many references.
Here we give a brief treatment for scalar fields in three dimensions.
A more general treatment including fields with spin may be found in \cite{Giombi:2008vd,David:2009xg}.

Consider 3-dimensional hyperbolic space ${\cal H}^3$ with metric
\be
ds^2_{{\cal H}^3} = d\rho^2 + \sinh^2 \rho \, d\Omega_2^2
\ee
We'll consider some slight generalizations below, but for now we'd like to determine the heat kernel for the Laplacian on this space.
\be
K_3(s,x,x') = \langle x \vert e^{-s\left(- \nabla_3^2\right)} \vert x' \rangle
\ee
Since ${\cal H}^3$ is maximally symmetric we can put $x'$ at the origin; the heat kernel will only depend on the radial coordinate of the other point.  For a field of mass $m^2$ the Green's function
is related to the heat kernel by
\be
G(x,x';m^2) = \langle x \vert {1 \over - \nabla_3^2 + m^2} \vert x' \rangle = \int_0^\infty ds \, e^{-s m^2} K_3(s,x,x')
\ee
It's straightforward to construct the Green's function by solving the radial differential equation
away from the origin and imposing the appropriate short-distance behavior.
\bea
\nonumber
&& \left(- {1 \over \sinh^2 \rho} \partial_\rho \sinh^2 \rho \partial_\rho + m^2\right) G(\rho,0;m^2) = 0 \qquad \hbox{\rm for $\rho \not= 0$} \\
&& G(\rho,0;m^2) \sim {1 \over 4 \pi \rho} \qquad \hbox{\rm as $\rho \rightarrow 0$}
\eea
This leads to
\be
G(\rho,0;m^2) = {e^{-\rho\sqrt{m^2+1}} \over 4 \pi \sinh \rho}
\ee
The heat kernel is given by an inverse Laplace transform.
\be
K_3(s,x,x') = \int_{c-i\infty}^{c+i\infty} {dm^2 \over 2 \pi i} e^{sm^2} G(x,x';m^2)
\ee
where $c > -1$ so that the contour runs vertically to the right of all singularities.  The contour can be deformed to enclose the branch cut at $-\infty < m^2 < -1$.  Integrating the discontinuity across the
cut leads to
\be
K_3(s,\rho,0) = \langle \rho \vert e^{-s\left(- \nabla_3^2\right)} \vert 0 \rangle = {\rho \, e^{-s} \, e^{-\rho^2 / 4 s}\over (4 \pi s)^{3/2} \sinh \rho}
\ee
It's trivial to extend this result to include a mass term.
\be
K_3^{\rm mass}(s,\rho,0) = \langle \rho \vert e^{-s\left(- \nabla_3^2 + m^2 \right)} \vert 0 \rangle = {\rho \, e^{-s(m^2 + 1)} \, e^{-\rho^2 / 4 s}\over (4 \pi s)^{3/2} \sinh \rho}
\ee
This includes the possibility of a non-minimal coupling which behaves as a mass term with $m^2 = \xi {\cal R}$.  Somewhat curiously conformal coupling in 4D means $m^2 = -1$ since
$\xi = {1 / 6}$ while the curvature of ${\cal H}^3$ is ${\cal R} = -6$.  Thus for conformal coupling
\be
K_3^{\rm conformal}(s,\rho,0) = {\rho \, e^{-\rho^2 / 4 s}\over (4 \pi s)^{3/2} \sinh \rho}
\ee
At coincident points we recover the result used in (\ref{replacement}).
\be
K_3^{\rm conformal}(s,x,x) = {1 \over (4 \pi s)^{3/2}}
\ee
More generally for a massive scalar at coincident points we have
\be
\label{MasslessScalarK}
K_3^{\rm mass}(s,x,x) = {e^{-s(m^2 + 1)} \over (4 \pi s)^{3/2}}
\ee
The massless limit of this result was used in (\ref{HeatKernels}).

\providecommand{\href}[2]{#2}\begingroup\raggedright\endgroup

\end{document}